\documentclass[prb,aps,showpacs,twocolumn]{revtex4}
\usepackage{amssymb}
\usepackage{amsmath}
\usepackage{graphicx}

\input{tcilatex}

\begin{document}

\title{\textbf{A single photon produces general W state of N qubits and its
application}}
\author{Di Mei}
\author{Chong Li }
\author{He-Shan Song}\email{hssong@dlut.edu.cn}
\affiliation{Department of Physics, Dalian University of
Techology, Dalian 116024, P. R. China }

\begin{abstract}
Based on the Wu's scheme[1], We prepare the general N-qubit W
state. We find that the concurrence of two qubits in general
N-qubit W state is only related to their coefficients and we
successfully apply the general N-qubit W state to quantum state
transfer and quantum state prepare like that in two-qubit system.
\end{abstract}

\maketitle

\section{Introduction}

Quantum entanglement plays an important role in quantum
information theory (QIT). It is regarded as an important resource
in various kinds of quantum information processing and quantum
computation[2]. Recently, there has been characterizing the
entanglement properties of multi-particle systems. Several parties
being spatially separated from each other and sharing a composite
system of an entangled state buildup a situation of particular
which is interest in QIT. This setting requires communication
through a classical channel and only act locally on their
subsystems. Although, the parties can still modify the
entanglement properties of the system and try to convert one
entangled state into another, even being restricted to local
operations assisted with classical communication[3]. \newline

The definition of W state was firstly given by W. D\"{u}r. G.
Vidal. and J. I. Cirac[4]. In that paper, W state consists of
three qubits, and the
coefficients are the same,%
\begin{equation*}
|W\rangle =\frac{1}{\sqrt{3}}(|001\rangle +|010\rangle
+|100\rangle )
\end{equation*}%
Here, we give a general W state of N-qubit that is defined as
\begin{equation*}
|W\rangle =\sum_{i}c_{i}|i\rangle ,|i\rangle =|0_{1}0_{2}\ldots
1_{i}\ldots 0_{N}\rangle
\end{equation*}%
where$|c_{1}|^{2}+|c_{2}|^{2}+|c_{3}|^{2}+\ldots
+|c_{N}|^{2}=1,$not all of the $c_{i}$ are equalities$.$ In
Sec.II, we find that the concurrence of
arbitrary m,n qubit in this state is only related to $|c_{m}|^{2}$ and $%
|c_{n}|^{2}$, and we give a new form of total concurrence of
general N-qubit W state. By contrasting quantum state transfer and
quantum state prepare in two-qubit system and general N-qubit W
state, It suggests that we could transmit the quantum information
by general W state as the same as that in two particles system.

In Ref.[1], the device gives us a way to generate W state, and by
this way, we can prepare the general N-qubit W state. In that
device, the teleportation is accomplished with mode entanglement.
Mode entanglement recently becomes more and more useful in the
quantum state transmit(QST)[5],[6],[7].

But for a single photon mode, we couldn't measure the photon to
make sure its state if no capturing the photon. This leads that we
have no way to realize the quantum information transmission. So we
make some change about the device in order to capture the photon
and finish the teleportation.

\section{concurrence of general N-qubit W state}

Usually, concurrence is widely used to measure pair-wise
entanglement defined in spin-1/2 systems[8], it is the measurement
of the two qubits entanglement[9],[10],[11]. The larger of
concurrence, the more powerful of the entanglement is to be. Two
qubits concurrence computation is given in
Ref.[12], $\rho _{ab}$ is the density matrix of system. Then we give the $%
\widetilde{\rho }_{ab}:$

\begin{equation}
\widetilde{\rho }_{ab}=(\sigma _{y}\otimes \sigma _{y})\rho
_{ab}^{\ast }(\sigma _{y}\otimes \sigma _{y})
\end{equation}

We define $R=\rho _{ab}\widetilde{\rho }_{ab}$, and the
eigenvalues of $R$ that are $\lambda _{i},i=1,2,3,4$.

As given in Ref.[12], the concurrence of the system is

\begin{eqnarray}
C_{ab} &=&\max \{0,\sqrt{\lambda _{1}}-\sqrt{\lambda _{2}}-\sqrt{\lambda _{3}%
}-\sqrt{\lambda _{4}}\} \\
&&\lambda _{1}\geq \lambda _{2}\geq \lambda _{3}\geq \lambda _{4}
\notag
\end{eqnarray}%
\ \

Here, we calculate the concurrence of general N-qubit W state. The
density matrix of general W state is given as follows.
\begin{equation}
\rho =|W\rangle \langle W|=\sum_{i,j}c_{i}^{\ast }c_{j}|j\rangle
\langle i|
\end{equation}%
$|i\rangle =|0_{1}0_{2}\ldots 1_{i}\ldots 0_{N}\rangle $.

For getting the density matrix of the m-th and the n-th qubit, we
take the trace of other qubits except the m-th,n-th qubit.
\begin{eqnarray}
\rho _{m,n} &=&\sum\limits_{k^{%
{\acute{}}%
}\neq m,n}\left\langle k^{%
{\acute{}}%
}\right\vert \rho \left\vert k^{%
{\acute{}}%
}\right\rangle  \notag \\
&=&\left(
\begin{array}{cccc}
1-|c_{m}|^{2}-|c_{n}|^{2} & 0 & 0 & 0 \\
0 & |c_{n}|^{2} & c_{m}^{\ast }c_{n} & 0 \\
0 & c_{n}^{\ast }c_{m} & |c_{m}|^{2} & 0 \\
0 & 0 & 0 & 0%
\end{array}%
\right)
\end{eqnarray}%
where the basis $\left\vert k^{%
{\acute{}}%
}\right\rangle =|0_{1}0_{2}\ldots 0_{n-1}0_{n+1}\ldots
0_{k-1}1_{k}0_{k+1}\ldots 0_{m-1}0_{m+1}\ldots 0_{N}\rangle $

Easily, we get the eigenvalues of $R_{m,n}$, and from equ.3, we
have the concurrence between m and n qubit.
\begin{equation}
C_{m,n}=2|c_{m}||c_{n}|
\end{equation}

\bigskip And as the similar form to it in Ref.[12] the total concurrence of
general N-qubit W state is%
\begin{equation*}
C_{total}=\sqrt{\frac{1}{2}\sum_{n\neq m}\left\vert C_{m,n}\right\vert ^{2}}=%
\sqrt{2\sum_{n\neq m}|c_{m}|^{2}|c_{n}|^{2}}
\end{equation*}

For this form, we find that when $|c_{1}|=|c_{2}|=\ldots =|c_{N}|=\frac{1}{%
\sqrt{N}},$ total concurrence reaches its maximum and
$C_{total_{\max }}\geq 1$. But we hope the maximal total
concurrence to be 1. So we give a new form
of total concurrence:%
\begin{equation}
C_{total}=\frac{\sqrt{2\sum_{n\neq m}|c_{m}|^{2}|c_{n}|^{2}}}{\sqrt{2(1-%
\frac{1}{N})}}=\sqrt{(\frac{N}{N-1})\sum_{n\neq
m}|c_{m}|^{2}|c_{n}|^{2}}
\end{equation}%
where N is the number of qubit, $C_{total}\leq 1.$ And when $%
N=2,|c_{1}|^{2}=|c_{2}|^{2}$, we will easily find the concurrence
of ours is the same as Wootters concurrence.

As mentioned in Ref.[5], we write another form of total
concurrence by using
equ.5.%
\begin{equation}
C_{total}=\sum_{j}^{N/2}C_{j,N+1-j}=2\sum_{j}^{N/2}|c_{j}||c_{N+1-j}|
\end{equation}%
or

\begin{equation}
\text{ }C_{total}=2\sum_{j}^{(N+1)/2}|c_{j}||c_{N+1-j}|
\end{equation}%
For equ.7, N is even, and N is odd number in equ.8.

In Ref.[5], quantum state transfer is described as a dynamic
process that starts from a initially factorized state to a
factorized finally state though an entanglement state in the
intermediate process. In that paper, when the fidelity reach
maximum, the total concurrence approximately vanishes. And the
maximal concurrence is 1. Here, we try to find the relationship
between two kinds of concurrence. In our paper, $C_{total}=0$,
when $|c_{i}|\neq 0$ and $|c_{k\neq i}|=0$ or $C_{j,N+1-j}=0$. When $%
|c_{i}|\neq 0$ and $|c_{k\neq i}|=0,$ it means that we only have
one state, necessarily the fidelity is 1. If $C_{j,N+1-j}=0$, the
system could not
accomplish quantum state transfer. In Ref.[5], $C(t)=\sum C_{j,N+1-j}$, if $%
C_{j,N+1-j}=0,$ they couldn%
\'{}%
t take the mode entanglement as the intermediate state. So the
processing of QST is transfer quantum state from $|1\rangle $ to
$|N\rangle $ through the middle state $|\psi (t)\rangle
=U(t)|1\rangle $. It also finishes the QST by
one state. Obviously, we have the same result. Then we find that when $%
|c_{1}|=|c_{2}|=\ldots =|c_{N}|=\frac{1}{\sqrt{N}}$, we receive $%
C_{total_{\max }}=1$. Therefor, we get the same result again. Let%
\'{}%
s review that paper, it suggests that $C(t)$ only refers to the
pair qubits
of mirror symmetry, the form is suitable for N as even. It implies that $%
C(t) $ is a especial kind of $C_{total}$. Whereupon, we explain
the processing of QST in Ref.[5] by our way. Because of more
generalization, we could consider the mode mentioned in Ref.[5] as
a especial condition or a part of ours.

\section{the application of general N-qubit W state}

\bigskip From Ref.[12], we know that for a two-qubit system $\left\vert \psi
\right\rangle =c_{1}\left\vert 01\right\rangle +c_{2}\left\vert
10\right\rangle $, we have
\begin{equation}
C_{total}=2|c_{1}||c_{2}|
\end{equation}%
There is surprisingly similarity between equ.5 and equ.9. So when
Alice
wants to transfer quantum state from her qubit to Bob%
\'{}%
s qubit, if they do not know their qubits in a two-qubit system or
in the general N-qubit W state, we want to know whether there will
be some difference. First, Alice transferes a unknown quantum
state $|\varphi (0)\rangle =\alpha |0\rangle +\beta |1\rangle $
with two different system. In two-qubit system $\left\vert \psi
\right\rangle =c_{1}\left\vert 01\right\rangle +c_{2}\left\vert
10\right\rangle $, after Bell combined-measurement, subsidiary
particle introduction and unitary operation, we find that when
$c_{1}>c_{2}$ , we have the probability which
is $2|c_{2}|^{2}$ , oppositely, the probability is $2|c_{1}|^{2}$ , only as $%
c_{1}=c_{2},$ we get the maximal probability.

In general N-qubit W state
\begin{equation*}
|W\rangle =c_{1}|00\cdots 1\rangle +c_{2}|0\cdots 10\rangle
+\cdots +c_{N}|10\cdots 00\rangle
\end{equation*}

The whole state is
\begin{eqnarray}
|\Phi \rangle &=&|\varphi (0)\rangle |W\rangle  \notag \\
&=&c_{1}\alpha |000\cdots 1\rangle +c_{1}\beta |100\cdots 1\rangle  \notag \\
&&+c_{2}\alpha |00\cdots 10\rangle +c_{2}\beta |10\cdots 10\rangle
+\cdots
\notag \\
&&+c_{N}\alpha |010\cdots 00\rangle +c_{N}\beta |110\cdots
00\rangle
\end{eqnarray}

Alice operates the C-not on the first two qubits, she obtains
that:
\begin{eqnarray}
&&c_{1}\alpha |000\cdots 1\rangle +c_{1}\beta |110\cdots 1\rangle
+c_{2}\alpha |00\cdots 10\rangle +  \notag \\
&&c_{2}\beta |110\cdots 10\rangle +\cdots +c_{N}\alpha |010\cdots
00\rangle +c_{N}\beta |100\cdots 00\rangle  \notag
\end{eqnarray}

Then she does the Hadmard operation on the first qubit, the result
is
\begin{eqnarray}
&&\frac{1}{\sqrt{2}}|00\rangle (c_{1}\alpha |00\cdots 1\rangle
+c_{2}\alpha
|00\cdots 10\rangle  \notag \\
&&+c_{2}\alpha |00\cdots 100\rangle +\cdots +c_{N}\beta |00\cdots
00\rangle
)+  \notag \\
&&\frac{1}{\sqrt{2}}|10\rangle (c_{1}\alpha |00\cdots 1\rangle
+c_{2}\alpha
|00\cdots 10\rangle  \notag \\
&&+c_{2}\alpha |00\cdots 100\rangle +\cdots -c_{N}\beta |00\cdots
00\rangle
)+  \notag \\
&&\frac{1}{\sqrt{2}}|01\rangle (c_{1}\beta |00\cdots 1\rangle
+c_{2}\beta
|00\cdots 10\rangle  \notag \\
&&+c_{2}\beta |00\cdots 100\rangle +\cdots +c_{N}\alpha |00\cdots
00\rangle
)+  \notag \\
&&\frac{1}{\sqrt{2}}|11\rangle (c_{N}\alpha |00\cdots 00\rangle
-c_{1}\beta
|00\cdots 1\rangle  \notag \\
&&-c_{2}\beta |00\cdots 10\rangle -c_{N-1}\beta |10\cdots
000\rangle ) \notag
\end{eqnarray}%
\ \

We take $\frac{1}{\sqrt{2}}|00\rangle (c_{1}\alpha |00\cdots
1\rangle +c_{2}\alpha |00\cdots 10\rangle +c_{2}\alpha |00\cdots
100\rangle +\cdots +c_{N}\beta |00\cdots 00\rangle )$ for
example.\newline

For $c_{1}\alpha |00\cdots 1\rangle +c_{2}\alpha |00\cdots
10\rangle +c_{2}\alpha |00\cdots 100\rangle +\cdots +c_{N}\beta
|00\cdots 00\rangle $ , after measuring by $\left\langle
0_{3}0_{4}\cdots 0_{N-1}\right\vert $, Bob has
$|c_{1}|^{2}+|c_{N}|^{2}$ probability to get the $|\varphi
(0)=\alpha |0\rangle +\beta |1\rangle $. He gets the total rate is $\frac{1}{%
4}(|c_{1}|^{2}+|c_{N}|^{2})$ . He wants to get the maximal
probability. So He works the $\hat{U}$ on the $c_{1}\alpha
|00\cdots 1\rangle +c_{2}\alpha |00\cdots 10\rangle +c_{2}\alpha
|00\cdots 100\rangle +\cdots +c_{N}\beta |00\cdots 00\rangle $.
$\hat{U}$ is a $M\times M$ matrix, $M=2^{N-1}$. The elements of
matrix of $\hat{U}$ is
\begin{eqnarray*}
\text{ }\hat{U}_{2,2} &=&t\text{ , }\hat{U}_{3,M}=t^{\ast }\text{ } \\
\hat{U}_{3,2} &=&-\sqrt{(1-|t|^{2})}\text{ , }\hat{U}_{2,M}=\sqrt{(1-|t|^{2})%
} \\
\hat{U}_{k,k} &=&1,k\neq 2,3,M\text{, }\hat{U}_{M,L}=1,L=2^{N-2}+1
\end{eqnarray*}%
Other $\hat{U}_{i,j}=0$
\begin{eqnarray}
t &=&\frac{1}{2}\{1-(-1)^{\frac{3}{2}[\frac{%
|c_{1}^{2}-c_{N}^{2}|+(c_{1}^{2}-c_{N}^{2})}{c_{1}^{2}-c_{N}^{2}}]}\}\frac{%
c_{N}}{c_{1}}  \notag \\
&&+\frac{1}{2}\{1-(-1)^{\frac{3}{2}[\frac{%
|c_{N}^{2}-c_{1}^{2}|+(c_{N}^{2}-c_{1}^{2})}{c_{N}^{2}-c_{1}^{2}}]}\}\frac{%
c_{1}}{c_{N}}  \notag
\end{eqnarray}%
Here, we calculate the processing of quantum state transfer from
the first qubit to the last. By similar computation, the result
suggest that we also could realize transfer from n qubit to m
qubit in this way.

After operation using $\left\langle 0_{3}0_{4}\cdots
0_{N-1}\right\vert $ , we find that when $c_{1}>c_{N}$ , we have
the probability of $2|c_{N}|^{2}$ , oppositely, the probability is
$2|c_{1}|^{2}$ , only as $c_{1}=c_{N},$ we get the maximal
probability. We generalize the computation to the W state like
that mentioned in Ref.[4]. By analogy, we get the maximal
probability that is $\frac{2}{N}$.

Above all, we find that there are no different results between two
systems when they are used in quantum state transfer.

As the same process, we accomplish quantum state prepare in two
systems, a known state$\left\vert \phi \right\rangle =\alpha
\left\vert 0\right\rangle +\beta \left\vert 1\right\rangle $
$\alpha ,\beta $ are real and we give a normal state $\beta
\left\vert 0\right\rangle -\alpha \left\vert 1\right\rangle
=\left\vert \psi \right\rangle $, then we express $\left\vert
0\right\rangle $ and $\left\vert 1\right\rangle $ with $\left\vert
\phi \right\rangle $ and $\left\vert \psi \right\rangle $.

\begin{eqnarray}
\left\vert 0\right\rangle &=&\alpha \left\vert \phi \right\rangle
+\beta
\left\vert \psi \right\rangle \\
\left\vert 1\right\rangle &=&-\alpha \left\vert \psi \right\rangle
+\beta \left\vert \phi \right\rangle
\end{eqnarray}

For $\left\vert \psi \right\rangle =c_{1}\left\vert
01\right\rangle +c_{2}\left\vert 10\right\rangle $, we obtain

\begin{eqnarray}
&&c_{1}\left\vert 01\right\rangle +c_{2}\left\vert 10\right\rangle \\
&=&\left( c_{2}\beta ^{2}-c_{1}\alpha ^{2}\right) \left\vert \phi
\right\rangle _{1}\left\vert \psi \right\rangle _{2}+\left(
c_{1}\beta ^{2}-c_{2}\alpha ^{2}\right) \left\vert \psi
\right\rangle _{1}\left\vert
\phi \right\rangle _{2}  \notag \\
&&+\left( c_{1}+c_{2}\right) \alpha \beta \left\vert \phi
\right\rangle _{1}\left\vert \phi \right\rangle _{2}-\left(
c_{1}+c_{2}\right) \alpha \beta \left\vert \psi \right\rangle
_{1}\left\vert \psi \right\rangle _{2} \notag
\end{eqnarray}%
when $c_{1}=-c_{2}$, we make the measurement by $_{1}\left\langle
\phi \right\vert $ and then finish the quantum state prepare.

For $|W\rangle =c_{1}|00\cdots 1\rangle +c_{2}|0\cdots 10\rangle
+\cdots
+c_{N}|10\cdots 00\rangle $. We replace $\left\vert 0\right\rangle $ and $%
\left\vert 1\right\rangle $ with $\left\vert \phi \right\rangle $ and $%
\left\vert \psi \right\rangle $ on the first and the last qubit.
We have the equation that is

\begin{eqnarray}
&&c_{1}|00\cdots 1\rangle +c_{2}|0\cdots 10\rangle +\cdots
+c_{N}|10\cdots
00\rangle  \notag \\
&=&\left\vert 0_{2}0_{3}0_{4}\cdots 0_{N-1}\right\rangle \{\left[
c_{N}\beta ^{2}-c_{1}\alpha ^{2}\right] \left\vert \phi
_{1}\right\rangle \left\vert
\psi _{N}\right\rangle +  \notag \\
&&\left[ c_{1}\beta ^{2}-c_{N}\alpha ^{2}\right] \left\vert \psi
_{1}\right\rangle \left\vert \phi _{N}\right\rangle +\left(
c_{1}+c_{N}\right) \beta \alpha \left\vert \phi _{1}\right\rangle
\left\vert
\phi _{N}\right\rangle  \notag \\
&&-\left( c_{1}+c_{N}\right) \alpha \beta \left\vert \psi
_{1}\right\rangle \left\vert \psi _{N}\right\rangle
\}+c_{2}\left\vert 1_{2}0_{3}0_{4}\cdots
0_{N-1}\right\rangle  \notag \\
&&\left( \alpha \left\vert \phi _{1}\right\rangle +\beta
\left\vert \psi _{1}\right\rangle \right) \left( \alpha \left\vert
\phi _{N}\right\rangle
+\beta \left\vert \psi _{N}\right\rangle \right) +  \notag \\
&&c_{3}\left\vert 0_{2}1_{3}0_{4}\cdots 0_{N-1}\right\rangle
\left( \alpha \left\vert \phi _{1}\right\rangle +\beta \left\vert
\psi _{1}\right\rangle \right) (\alpha \left\vert \phi
_{N}\right\rangle +\beta \left\vert \psi
_{N}\right\rangle )  \notag \\
&&+\cdots +c_{N-1}\left\vert 0_{2}0_{3}0_{4}\cdots
1_{N-1}\right\rangle \left( \alpha \left\vert \phi
_{1}\right\rangle +\beta \left\vert \psi
_{1}\right\rangle \right)  \notag \\
&&\left( \alpha \left\vert \phi _{N}\right\rangle +\beta
\left\vert \psi _{N}\right\rangle \right)
\end{eqnarray}

After measuring the second qubit with $\left\langle
0_{2}0_{3}0_{4}\cdots 0_{N-1}\right\vert $, we\ find that\ if
$c_{1}=-c_{N}$, we could easily accomplish the quantum state
prepare. And the probability is $2|c_{1}|^{2}$. Obviously, there
are also the same between two systems in quantum state prepare.

Above results indicate that we could apply the general N-qubit W
state to the quantum state transfer and quantum state prepare like
that in two-qubit system, we could use the same way to get the
similar result.

\section{a scheme to prepare general N-qubit W state and QST}

The device given in Ref.[1] could generate W state that has the
same probability. As Fig.1, We find that we could change the
reflectivity of the each beam splitter to get the general W state.
Here we use the same probability W state for convenience and no
losing general.

There is a single photon to prepare W state by mode entanglement
or accomplish quantum sate transfer in the schemes in
Refs.[1],[13],[14].We agree that the prepare of W state is viable,
but there are two difficulties in the state transfer. One is the
question of photon settlement. Without staying photon, we will not
measure on the photon to realize the quantum state transfer for
too long distance of photon having escaped. The other question
comes from mode entanglement. In single photon system, the being
or not of photon consist of the mode entanglement. Once we measure
the photon, the W state will collapse to a single state. So we
couldn't measure the qubit in this condition, then we are no way
to transmit the quantum information. Therefor, we need try to make
some change about the device in order to realize the measurement.
In this view, we take some cavities that they can capture the
photon instead of photon counting detectors. In this way, we could
finish distribute these cavities like particles. The people who
have the cavities don't know whether there is photon in them. The
new W state is $|W\rangle =c_{1}|1_{1}0_{2}\ldots 0_{N}\rangle
+c_{2}|0_{1}1_{2}\ldots 0_{N}\rangle +\ldots
+c_{N}|0_{1}0_{2}\ldots 1_{N}\rangle $. Here $|1_{i}\rangle $ or
$|0_{i}\rangle $indicates that there is a photon or not in the i
position hollow. After distributed, put a particle, unknown its
state, into the first hollow. Then we make the operation on the
cavities like doing on the particles, and we will get the state in
some probability. With this changed device, we realize a single
photon to prepare general N qubits W state and transmit the
quantum information.

\section{conclusion}

We have obtained the general W state by the device as Fig1. In
this mode, we calculate the concurrence of the general N-qubit W
state. As the result, the $C_{m,n}$ is only related to $|c_{m}|$
and $\left\vert c_{n}\right\vert $, and give a new form of total
concurrence of general N-qubit W state. Then by contrasting
quantum state transfer and quantum state prepare in two-qubit
system and general N-qubit W state, we find we will get the same
result whichever system the two qubits that transfer quantum state
from one to another in. At last we improve the device to give a
scheme that realize quantum state transfer and quantum state
prepare. In this paper, we make the mode entanglement be useful in
the processing of QST.

\begin{acknowledgments}
This work was partially supported by the CNSF (grant No.10575017,
and No.10547106) .
\end{acknowledgments}

\end{document}